\documentstyle[sprocl,epsf]{article}
\begin{document}

\newcommand{\yp}{y}
\newcommand{\z}{&\hspace*{-8pt}}
\newcommand{\eps}{\varepsilon}
\newcommand{\logtwos}{L_{tW}}
\newcommand{\logtwms}{l_{tW}}
\newcommand{\logctheta}{l_\Theta}
\newcommand{\logmuW}{l_{\mu W}}

\font\elevenit=cmti10 scaled\magstephalf

\begin{flushright}
\end{flushright}

\begin{center}


{\Large \bf Techniques for Calculating two-loop Diagrams}

\vskip 10mm

J.~Fleischer
\footnote{~E-mail: fleischer@physik.uni-bielefeld.de}
and
O.~L.~Veretin
\footnote{~E-mail:veretin@ifh.de; present address: DESY-Zeuthen, Platanenallee 6,
D-15738 Zeuthen, Germany}
\footnote{~Supported by BMBF under 05~7BI92P(9)}

\vskip 10mm

{\it ~Fakult\"at f\"ur Physik, Universit\"at Bielefeld,
D-33615 Bielefeld, Germany.}

\begin{abstract}

  Methods developed by the Bielefeld-DESY-Dubna collaboration in recent
  years are:~DIANA (DIagram ANAlyser), a program to produce ``FORM input''
for Feynman diagrams,
  starting from the Feynman rules; methods to calculate scalar diagrams:
Taylor expansion in small momenta squared in connection with a mapping
and the Pad\'{e} method to sum the series. Recently program packages for
the large mass expansion were written and applied to the
$Z \to b\bar{b}$ decay. Reviews of these activities were presented in
the proceedings of the Ustro\'{n} '97 and Rheinsberg '98 conferences.
Here we concentrate on recent developements in the large mass expansion,
applied to the two-loop contribution of the $Z \to b\bar{b}$ decay
in the $m_b=0$ approximation, taking into account higher order terms
of the expansion in $M_W^2/m_t^2$.

\end{abstract}
\end{center}


\thispagestyle{empty}
\setcounter{page}0

\section{Introduction}

   The calculation of diagrams with one non-zero external momentum squared 
($q^2$) has wide applications in QED and QCD for both selfenergies and
vertices. In these cases also only one non-zero mass enters
the problem. In electroweak problems like $Z \to b\bar{b}$ one has
mixing terms between electroweak and strong interactions and due to
that different internal masses occur, so that the method of Taylor expansion
is getting more difficult to apply. In this case,
however, the top quark ($m_t$) plays a special role and it
allows to make the
expansion in the large mass. The method is not applicable to
arbitrary high $q^2$, but as has been demonstrated in \cite{LM}, for
$q^2=m_Z^2$ this approach is still reliable. While in \cite{LM} only
scalar diagrams have been considerd, here we investigate the full
decay amplitude. It turns out that the obtained results are simpler
for the full process than for scalar diagrams in the following sense:
first of all, complicated functions like higher polylogarithms, which
show up in the analytic evaluation of scalar diagrams, cancel in the
full amplitude; furthermore the convergence of the large mass
expansion also turns out to be better for the full amplitude than for
scalar diagrams. These observations are manifestation
of gauge cancellations observed in gauge theories in general.
Nevertheless their observation in this special form is surprising!

  Due to the fact that the  method of mapping and Taylor expansion is
  quite useful and finds applications by other authors (see e.g.
  \cite{Kue}), we give a
  short review here concerning the method and report on recent
  developements of
  calculating Taylor coefficients for two-loop diagrams.
  Then, in the second part, we turn to our main point 
  namely the large mass expansion for the $Z \to b\bar{b}$ problem
  and the comparison with the work of \cite{HSS}.

\section{Expansion of three-point functions in terms of an
external momentum squared}

      Taylor series expansions in terms of one external momentum
squared, $q^2$ say, were considered for selfenergy diagrams
in \cite{SE}, Pad\'{e} approximants
were introduced in \cite{bft} and in Ref. \cite{ft} it was demonstrated
that this approach can be used to calculate Feynman diagrams on their
cut by analytic continuation.
In the case of a three-point function like $Z \to b\bar{b}$
in the $m_b=0$ limit we have for the external $b-$quark momenta
$p^2_1 = p^2_2 = 0$. The expansion of the scalar diagram then looks like

\begin{equation}
\label{eq:exptri}
C(p_1, p_2) = \sum^\infty_{n=0} a_{n} (p_1 p_2)^n,
\label{2.2}
\end{equation}
with $q^2=(p_1 + p_2)^2.$

For the calculation of the Taylor coefficients in general various procedures have
been proposed \cite{ZiF,davt,Tara}. 
These methods are well suited for programming in terms of a formulae manipulating 
language like FORM \cite{FORM}. Such programs, however, yield acceptable analytic results 
only in cases when not too many parameters (like masses) enter the problem. 
Otherwise numerical methods are needed \cite{JF}.

In the case of only one non-zero mass and only one external momentum squared,
indeed the case with the least nontrivial parameters, for many diagrams 
analytic expressions for the Taylor coefficients can be obtained. For
recent references see \cite{one}.

For the purpose of calculating Feynman diagrams in the kinematical domain
of interest it is necessary to calculate them from the Taylor series on 
their cut. This is performed by analytic continuation in terms of a mapping \cite{ft}

Assume, the following Taylor expansion of a scalar diagram or a
particular amplitude is given
$C(p_1, p_2,\dots)=\sum^\infty_{m=0} a_m y^m \equiv f(y)$
and the function on the r.h.s. has a cut for $y \ge y_0$.

 The method of evaluation
of the original series consists in a first step in a conformal mapping
of the cut plane into the unit circle and secondly the reexpansion
of the function under consideration
into a power series w.r.t. the new conformal variable.
We use
\begin{equation}
\omega=\frac{1-\sqrt{1-y/y_0}}{1+\sqrt{1-y/y_0}}.
\label{omga}
\end{equation}

\begin{figure}[h]
\centerline{\vbox{\epsfysize=45mm \epsfbox{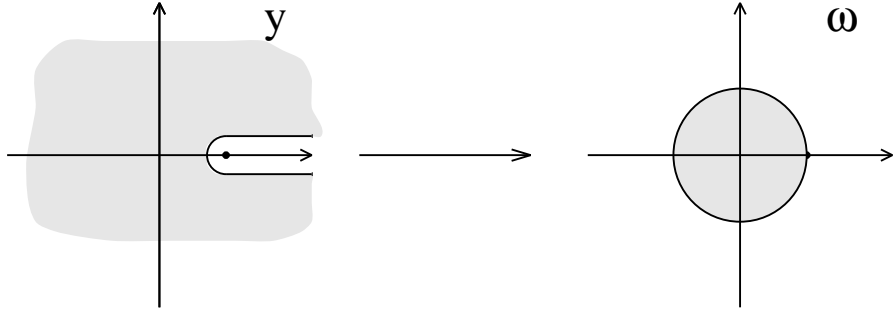}}}
\caption{\label{conf}Conformal mapping of the
complex y-plane into the $\omega$-plane.}
\end{figure}

By this conformal transformation,
the $y$-plane, cut from $y_0$ to $+ \infty$, is mapped into the unit
circle (see Fig.\ref{conf}) and the cut itself is mapped on
its boundary, the upper
semicircle corresponding to the upper side of the cut.
The origin goes into the point $\omega=0$.

  After conformal transformation it is suggestive to improve the
convergence of the new series w.r.t. $\omega$ by applying the
Pad\'e method \cite{Pade,Sha}.
A convenient technique for the evaluation of Pad\'e approximations
is the $\varepsilon$-algorithm of~\cite{Sha} which allows one
to evaluate the Pad\'e approximants recursively.

   Generally speaking, the precision of results with this mapping and
   Pad\'e is of the order of 3-4 decimals with 30 Taylor coefficients
for timelike $q^2$ values a factor of approximately 100 times the 
lowest threshold value. For lower $q^2$ 
(a few times the threshold value) the precision
is of the order of 10 decimals in quite many cases. The precision worsens
near second nonzero thresholds. 
 
   As a final remark we mention that for diagrams with zero thresholds
new techniques have been developed. In fact terms of the form
$\ln^m(q^2)$ have to be factorized, where $m$ is the number of zero
thresholds of the diagram. The factors in front are then expanded in
terms of Taylor series \cite{one,fst}.

\section{Large Mass Expansion (LME)}

  As mentioned above, for the evaluation of diagrams with several different
masses, one of which being large (like the top mass  $m_t$),
we use the general method of asymptotic
expansion in large masses \cite{asymptotic}. For a given scalar
graph $G$ the expansion in large mass is given by the formula
\begin{equation}
F_G(q, M ,m, \varepsilon) \stackrel{M \to \infty}{\sim }
\sum_{\gamma} F_{G/\gamma}(q,m,\varepsilon) \circ
T_{q^{\gamma}, m^{\gamma}}
F_{\gamma}(q^{\gamma}, M ,m^{\gamma}, \varepsilon),
\label{Lama}
\end{equation}
\noindent
where $\gamma$'s are subgraphs involved in
the asymptotic expansion, $G/\gamma$ denotes shrinking of $\gamma$ to a
point; $F_{\gamma}$ is the Feynman integral corresponding to
$\gamma$; $ T_{q_{\gamma}, m_{\gamma}} $ is the Taylor operator
expanding the integrand in small masses $\{ m_{\gamma} \}$ and
external momenta $\{ q_{\gamma} \}$ of the subgraph $\gamma$
; $ \circ$ stands for the convolution of the subgraph expansion
with the integrand $F_{G/{\gamma}}$. The sum goes over all
subgraphs $\gamma$ which (a) contain all lines with large masses, and
(b) are one-particle irreducible w.r.t. light lines.

   For the $Z \to b\bar{b}$ decay we have $q^2 = M_Z^2$ for the 
on-shell $Z$'s. Fig.2 shows diagrams with two different masses
on virtual lines, one of which a top. 
$W$ and $Z$ are the gauge bosons with masses $M_W$ and $M_Z$, respectively;
$\phi$ is the charged would-be Goldstone boson (we use the Feynman gauge);
$t$ and $b$ are the t- and b-quarks. Fig. 3 shows subdiagrams needed
in the expansion (3).

\begin{figure}[bth]
\centerline{\vbox{\epsfysize=50mm \epsfbox{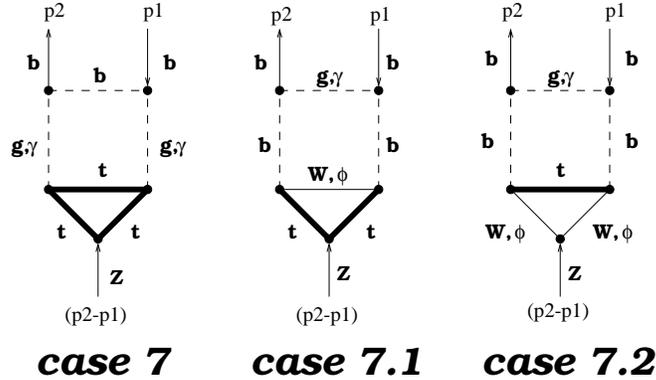}}}
\caption{\label{fig2} Two-loop diagrams with two different masses
on internal lines arising in $Z b \overline{b}$. }
\end{figure}

\begin{figure}[bth]
\centerline{\vbox{\epsfysize=50mm \epsfbox{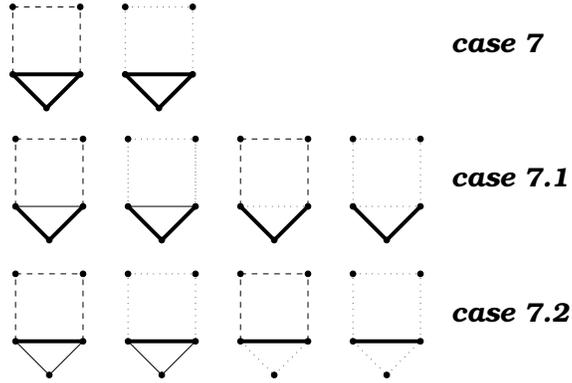}}}
\caption{\label{fig3} The structure of the LME, see explanations in the text. }
\end{figure}

Finally the LME of the above diagrams has the
following general form
\begin{equation}
F_{\rm as}^N = \frac{1}{m_t^4}
  \sum_{n=-1}^N \sum_{i,j=-1;i+j=n}^n 
  \left( \frac{M_W^2}{m_t^2} \right)^i \left( \frac{q^2}{m_t^2} \right)^j
  \sum_{k=0}^m A_{i,j,k}(q^2,M_W^2,{\mu}^2) \ln^k \frac{m_t^2}{\mu^2}
\label{series}
\end{equation}
where $m$ is the highest degree of divergence (ultraviolet, infrared, collinear)
in the various contributions to the LME ($m\le$ 3 in the 
cases considered). 
$M_W^2/m_t^2$ and $q^2/m_t^2$ are considered as small parameters. 
$A_{i,j,k}$ are in general complicated functions of the arguments, i.e. they
may contain logarithms and higher polylogarithms.

In contrary to the work of \cite{HSS} we see no inconveniences in
directly applying the above method and did not use any ``continued expansion''.

In the following we present some of our results of the LME for the full
two-loop $O(\alpha\alpha_s)$ contribution. 
As in \cite{HSS}, we are interested only in the
virtual effect of the top quark, which renders the decay of the $Z$
boson into bottom quarks different from the one into other down-type
quarks. Therefore in the following the quantity
$\Gamma_{b-d}^{W}=\Gamma_{Z\to bb}^{W}-\Gamma_{Z\to dd}^{W}$ is considered in 
two loop order, in which expression the counterterm contributions
cancel. The superscript $W$ means that only diagrams with virtual
$W$ bosons are included. The other part with the $Z$ exchange 
makes no discrimination between $b$- and $d$-quarks and is calculated in order
$\alpha\alpha_s$ in \cite{Zexchange}. Our result reads
\begin{eqnarray}
\lefteqn{ \delta \Gamma_{b-d}^{(2),W} = \Gamma^0
    {1\over s_{\Theta}^2}{\alpha\over \pi}\frac{\alpha_s}{\pi}  \times }
\nonumber\\&&\mbox{} 
\bigg\{
           {M_t^2\over M_W^2}\,\bigg[
               - {1\over 32} 
               - {1\over 64}\,{1\over \yp} 
               + \zeta_2\,\bigg(
                     {1\over 16} 
                   + {1\over 32}\,{1\over \yp}
                  \bigg)
              \bigg]
\nonumber\\&&\mbox{} 
           + \bigg[
               \frac{61}{288} \frac{1}{y} 
              + \frac{82661}{466560} 
              - \frac{106626671}{204120000}y
              + \frac{673933}{1458000} y^2 
              - \frac{12334491149}{16044682500}y^3
              + {\cal O} (\yp^4)
\nonumber\\&&\mbox{}
\hspace{1em}
           + I_1\,\biggl(  
                   \frac{1}{96} \frac{1}{y^2} 
                  - \frac{5}{192} \frac{1}{y} 
                  -\frac{5}{48} 
                  - \frac{1}{48}y 
                  \biggr)
\nonumber\\&&\mbox{}
\hspace{1em}
             + \zeta_2\,\bigg(
                  {173\over 1296} 
                  + {67\over 2592}\,{1\over \yp}
                  + {53\over 324}\,\yp 
                  \bigg) 
              + \zeta_3\,\bigg(
               - {1\over 18}\,{1\over \yp}
               + {7\over 90}\,\yp 
               - {1\over 45}\,\yp^{2} 
               - {16\over 315}\,\yp^3 
              \bigg)
\nonumber\\&&\mbox{}
\hspace{1em}
              + \logtwos\,\bigg(
                  - {757\over 7776} 
                  - {331\over 7776}\,{1\over \yp}
                  - {95\over 3888}\,\yp 
                  \bigg) 
              + \logctheta^2\,\bigg(
                  - {103\over 2592} 
                  - {1\over 81}\,{1\over \yp}
                  + {1\over 300}\,\yp 
\nonumber\\&&\mbox{}
\hspace{1em}
                  - {103\over 1080}\,\yp^2 
                  - {5314\over 33075}\,\yp^3 
                  \bigg) 
              + \logctheta\,\bigg(
                  - {527\over 7776} 
                  + {11\over 288}\,{1\over \yp}
                  - {1489\over 30375}\,\yp 
                  + {1081\over 9720}\,\yp^2 
\nonumber\\&&\mbox{}
\hspace{1em}
                  - {3338578\over 10418625}\,\yp^3 
                  \bigg)
              \bigg] 
\nonumber\\&&\mbox{}
           + {M_W^2\over M_t^2}\,\bigg[
              \frac{5}{128} \frac{1}{y} 
            - \frac{64847}{1555200} 
            - \frac{636239}{777600} y 
            - \frac{12497}{10800} y^2 
\nonumber\\&&\mbox{}
\hspace{1em}
            + I_1\,\biggl( 
              - \frac{11}{288} \frac{1}{y^2} 
              - \frac{253}{864} \frac{1}{y} 
              - \frac{11}{48} 
              + \frac{11}{24} y 
              + \frac{11}{108} y^2 
                    \biggr)
\nonumber\\&&\mbox{}
\hspace{1em}
               + \logtwos\,\bigg(
                   - {83083\over 155520} 
                   - {1819\over 5184}\,{1\over \yp}
                   + {15017\over 38880}\,\yp 
                   + {3977\over 38880}\,\yp^2 
                  \bigg) 
               + \logctheta\,\bigg(
                   - {3343\over 155520} 
                   - {7\over 1296}\,{1\over \yp}
\nonumber\\&&\mbox{}
\hspace{1em}
                   - {823\over 38880}\,\yp 
                   + {17\over 38880}\,\yp^2 
                  \bigg)
               + \zeta_2\,\bigg(
                     {257\over 864} 
                   + {175\over 864}\,{1\over \yp}
                   + {13\over 144}\,\yp 
                   + {11\over 18}\,\yp^2 
                  \bigg)
              \bigg], 
\label{eqzbbOS2}
\end{eqnarray}
where $\Gamma^0$ is the Born decay rate, $y=M_Z^2/4M_W^2$,
$l_\Theta=\ln\cos\Theta_W$, $L_{tW}=\ln(m_t^2/M_W^2)$
and $\zeta_n=\zeta(n)$ the Riemann $\zeta$-function.
The following integral is introduced
\begin{equation}
I_n = \frac12 \frac{(-1)^n}{n!} \int\limits_0^1
       \frac{{\ln^n(1-ty)}}{\sqrt{1-t}}\,dt\,.
\end{equation}

   In the above final result enters only $I_1$ which has the expansion
\begin{equation}
I_1 = \frac{2}{3}y 
         +\frac{4}{15}y^2
         +\frac{16}{105}y^3
         +\frac{32}{315}y^4
         +\frac{256}{3465}y^5
         +\frac{512}{9009}y^6
         +\frac{2048}{45045}y^7
         +\frac{4096}{109395}y^8
         + \dots
\end{equation}

   Inserting this expansion in (\ref{eqzbbOS2}), we fully agree with the result of 
Harlander et.al. \cite{HSS}, as far as they have presented their
result. Our result is more compact, however, and it is interesting 
to observe that,
while higher polylogarithms occur in the scalar integrals \cite{LM},
in the full decay amplitude these and the higher $I_n$ (also, however,
expressible in terms of polylogarithms) cancel. The remaining $I_1$,
expanded above, is merely a logarithm: $I_1=2-a\log\bigl((a+1)/(a-1)\bigr),
a=\sqrt{1-1/y}$. Thus we observe that the final result is much simpler
than intermediate results from scalar diagrams.
Moreover, the convergence of the series in terms of
large masses is much better than for the series obtained for
scalar diagrams \cite{LM}, which is demonstrated below.

Our numerical results are as follows: $x_1$ and $x_2$ being the 1-loop and
2-loop results, $r=M_W^2/m_t^2$, the large mass expansion is given in the form
\begin{eqnarray}
\delta\Gamma_{b-d}^{W} \z=\z 
  \Gamma^0 \frac{1}{s_\Theta^2} \frac{\alpha}{\pi}
  \Biggl[ x_1 + \frac{\alpha_s}{\pi} x_2 \Biggr] , \\
x_1 \z=\z 
  \Biggl(-\frac{0.1063}{r}\Biggr)
  + \Biggl(
    0.2360
  - 1.0236 r
  - 1.7492 r^2
  - 1.4477 r^3
  - 0.1758 r^4
  + 1.2997 r^5 \nonumber\\
  \z+\z
    2.1005 r^6
  + 1.7986 r^7
  + 0.5692 r^8
  - 0.9260 r^9
  - 1.6273 r^{10} \Biggr)\,,\\
x_2 \z=\z 
    \Biggl(\frac{0.2435}{r}\Biggr)
  + \Biggl(
    0.7167
  - 1.6940 r
  - 4.0920 r^2
  - 4.7543 r^3
  - 2.9143 r^4
  + 0.5688 r^5 \nonumber\\
\z+\z
    3.7747 r^6
  + 4.8688 r^7
  + 3.6932 r^8
  + 4.5196 r^9
  + 24.090 r^{10} \Biggr)\,.
\end{eqnarray}
The first terms in $x_1,x_2$ correspond to the leading $m_t^2/M_W^2$.
For $m_t=175$GeV, $M_W=80.33$GeV and $M_Z=91.187$GeV we obtain
\begin{equation}
\delta\Gamma^W_{b-d}
  = \Gamma^0 \frac{1}{s_\Theta^2} \frac{\alpha}{\pi}
  \Biggl[ -0.5045-0.0704 
     + \frac{\alpha_s}{\pi}\bigl( 1.1556+0.1285 \bigr) \Bigg]
\end{equation}
   In each of these terms the leading term and the corrections
are given separately. Note that the leading $m_t^2/M_W^2$
term in order $O(\alpha\alpha_s)$ was obtained earlier in \cite{ZbbF}.

  The observation is that the series for
the full amplitude converges like ~$r^n$ while for the scalar
diagram the convergence was like ~$(4r)^n$. Accordingly the
term of order $r^4$ gives only an error of order 0.5\% while
for the scalar diagrams the errors where of the order several \% \cite{LM}.

\end{document}